# Scattering-initiated parametric noise in optical parametric chirped-pulse amplification

Jing Wang,[1] Jingui Ma,[1] Peng Yuan,[1,*] Daolong Tang,[1] Binjie Zhou,[1] Guoqiang Xie,[1] and Liejia Qian[1,2]

[1]*Key Laboratory for Laser Plasma (Ministry of Education) and Department of Physics and Astronomy, IFSA Collaborative Innovation Center, Shanghai Jiao Tong University, Shanghai 200240, China*

[2]*e-mail: qianlj19@sjtu.edu.cn*

*Corresponding author: pengyuan@sjtu.edu.cn*

We experimentally study a new kind of parametric noise that is initiated from signal scattering and enhanced through optical parametric amplification. Such scattering noise behaves similarly to the parametric super-fluorescence in the spatial domain, yet is typically much stronger. In the time domain, it inherits the chirp of signal pulses and can be well compressed. We demonstrate that this scattering-initiated parametric noise has little influence on the amplified pulse contrast but can degrade the conversion efficiency substantially.

*OCIS Codes: (190.4410) Nonlinear optics, parametric processes; (190.4970) parametric oscillators and amplifiers.*

Optical parametric chirped-pulse amplification (OPCPA) has been the one workhorse in the race to develop extremely intense lasers [1-3]. Compared to chirped-pulse amplification (CPA) based on laser gain media, OPCPA has the combined advantages of high gain, broad bandwidth and negligible thermal load due to the use of parametric amplifiers. The availability of large-aperture nonlinear crystals and introduction of noncollinear phase-matching scheme then further enhance its superiority. Presently, advanced OPCPA systems producing 10-PW high-energy pulses and 1-PW few-cycle pulses have been under construction [4-6]. These lasers would extend strong field physics and ultrafast science to previously inaccessible regimes.

The fundamental limits on high-power laser applications, however, are often governed by the noise of optical amplifiers. In the context of OPCPA, investigations on the noise issue are primarily motivated by the requirement on pulse contrast of ultra-intense lasers in many applications. Previous studies have verified three kinds of noise in parametric amplifiers, including parametric super-fluorescence (PSF) [7-11], noise transferred from pump laser [12-13] and surface-reflection-initiated pre-pulses [14]. The PSF results from optical parametric generation (OPG) in high gain regime wherein the amplification of spontaneously emitted photons (i.e., quantum noise) becomes prominent. The latter two kinds of noise are both related to nonlinear transfer of intensity modulations via the instantaneous parametric gain. All of these three kinds of noise cannot be compressed like the chirped signal pulses, hence causing a dramatic degradation of pulse contrast. Besides these noise, OPCPA is also susceptible to various parasitic nonlinear processes, such as the frequency doubling of signal in degenerate OPCPAs [15]. Although such processes have little effect on pulse contrast, they could degrade the pump-to-signal conversion efficiency.

In this Letter, we report on the experimental observation of scattering-initiated parametric noise in an OPCPA. This is another kind of excess noise in parametric amplifiers besides PSF. Light scattering is an omnipresent physical process accompanying beam propagation, whatever it is molecular scattering in air path or diffuse scattering from the imperfections in optics [16]. Recent studies have verified that scattering in pulse stretchers and compressors is the origin of spatiotemporal noise in CPA systems [17-18]. Light scattering can also contaminate the single-shot measurement of pulse contrast [19]. However, the influence of scattering on parametric amplifiers has not been studied to date. Here we demonstrate that light scattering in the signal beam path will unavoidably introduce an initial noise field into parametric amplifiers, which then undergoes amplification similarly to the well-known PSF. Such scattering-initiated parametric noise (hereafter 'scattering noise' for short) is typically much stronger than PSF, and might be a predominant factor limiting the extractable signal energy for OPCPAs.

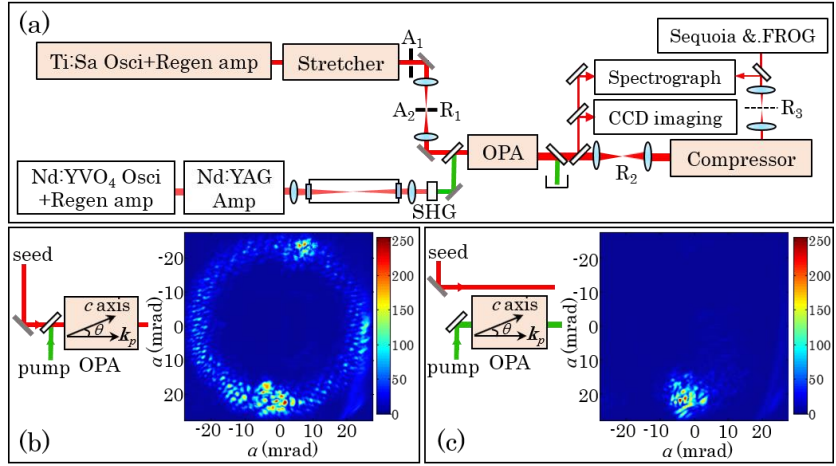

Fig. 1. (a) Schematics of the experimental setup. (b) CCD image of the conical scattering noise in the case that the signal injection is collinear with the pump. (c) CCD image of the scattering noise in the case that the seed beam is outside the OPA crystal.

Our experimental setup [Fig. 1(a)] is based on a single-stage type-I OPCPA using a 12-mm-thick $\beta$-BBO crystal. The pump laser system sequentially consists of a Nd:YVO$_4$ laser oscillator-regenerative amplifier (High-Q, Pico-Regen), a 10 Hz Nd:YAG boost amplifier system (Innolas, Spit Light) and a frequency doubling stage using a 5-mm-thick $\beta$-BBO crystal. This system provides 532 nm pump pulses with temporal duration of 420 ps and pulse energy of 90 mJ. The seed signal is produced by a commercial 1 kHz femtosecond Ti:Sapphire regenerative amplifier system (Coherent, Legend Elite), which delivers laser pulses centered at 800-nm with temporal duration of 35 fs. The synchronization between pump and signal (time jitter ~10 ps) was achieved by using an electronic phase-locking loop. The femtosecond signal pulses were then temporally stretched to 380 ps through a single-grating spherical Öffner stretcher. After that, a spatial filter ($R_1$) and two apertures ($A_1$ and $A_2$) were applied to clean up the signal beam in both the near field and far field. The beam path between aperture $A_2$ (0.5 mm in diameter) and the OPA crystal was exposed to the laboratory environment (class 1000 cleanroom). The signal injected into the OPA had a pulse energy of 90 $\mu$J and beam size of 2.8 mm. In order to preserve the amplified scattering noise, two additional telescopes ($R_2$ and $R_3$) were adopted to relay image the OPCPA output to the pulse compressor and diagnostic unit successively. The diagnostic unit consists of a spectrometer, a frequency-resolved optical gating (FROG) device (Femtosoft, GRENOUILLE 8-50 USB) and a third-order cross-correlator (Amplitude, SEQUOIA).

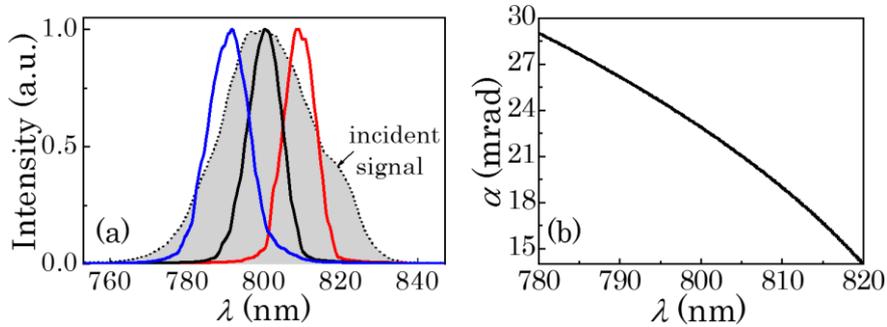

Fig. 2. (Color Online) (a) Spectra of the conical ring in Fig. 1(b) measured at $\alpha$ = 19.5 (red), 23 (black) and 26.5 (blue) mrad, respectively, and the spectrum of the incident signal (shadow). (b) Calculated phase-matching noncollinear angles for different seeding wavelengths.

To observe the scattering noise, we firstly set the parametric amplifier deviating from optimal phase-matching condition, wherein the signal pulses could not be effectively amplified. Specifically, the signal and pump beams were injected into the BBO crystal collinearly, while the crystal orientation was rotated from the collinear phase-matching angle $\theta$ = 22.06° to 22.26°. The pump intensity was fixed at $I_p$ = 2.0 GW/cm$^2$ (i.e., 51-mJ in energy) to provide a moderate small-signal gain of $1.7\times10^6$. In this case, the PSF (no signal injection) was as weak as 7.2 $\mu$J. With a signal injection of 90 $\mu$J, we observed a bright conical ring centered on the pump beam with an emission angle of $\alpha\approx$ 23 mrad, as presented in Fig. 1(b). This conical ring had an energy (0.6 mJ) significantly higher than the PSF and a global spectral extent identical to the incident signal pulses [Fig. 2(a)]. When we withdrew the signal injection, this conical ring disappeared entirely. These results consistently reflect that the observed conical ring was stemmed from the signal pulses. Since aperture $A_2$ had filtered out those high-spatial-frequency (>1.25 mrad) components carried by the signal beam, this conical ring was newly generated from the light scattering in the signal beam path after $A_2$. As an analogy, we point out that PSF can exhibit a similar conical ring pattern. When we blocked the signal injection and increased the pump intensity to 3.4

GW/cm$^2$ (parametric gain >10$^7$), PSF in the form of a solid cone covering a broad angular range can be clearly observed. After inserting an 800 nm band pass filter, this PSF cone reduced to a conical ring of ~23 mrad just like that in Fig. 1(b). We further confirmed that the scattering occurred mostly in the beam path before the crystal rather than that within the crystal. To this end, we deliberately shifted the signal injection to be totally out of the crystal. As shown in Fig. 1(c), we can still observe the scattering noise, but only the portion in the opposite side of signal beam was remained, with unchanged emission direction and equivalent brightness as the conical ring presented in Fig. 1(b). To ensure the experimental identification of scattering noise not confused by the weak beam edges of the signal, we have cleaned the incident signal beam in the near-field with aperture $A_1$.

Besides the similar spatial behavior of conical emission, the scattering noise acquires an angular dispersion during parametric amplification as like PSF [7]. Figure 2(a) presents the spectra of the conical ring in Fig. 1(b) measured along different emission directions $\alpha$. For a fixed $\alpha$, the conical ring had a local bandwidth of 11 nm. With the increase of $\alpha$ from 19.5 to 26.5 mrad, the central wavelength gradually shifted from 809 to 791 nm, indicating an angular dispersion of 0.39 mrad/nm. Figure 2(b) plots the theoretically calculated phase- matching non-collinear angle $\alpha$ for different seeding wavelengths in a BBO crystal tuned to $\theta = 22.26°$. This result shows a good agreement with our experimental measurement.

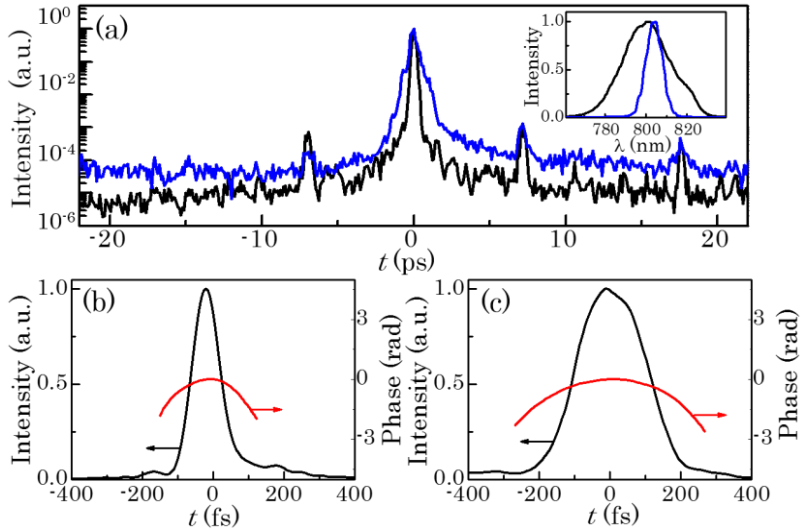

Fig. 3. (Color Online) (a) Third-order cross-correlation traces of the signal pulse (black) and the conical ring (blue) after passing through the same pulse compressor. Inset: the corresponding spectra measured after the compressor. (b), (c) Temporal profiles of the compressed signal pulse and scattering noise, respectively.

To study the scattering noise characteristics in the time domain, the OPCPA output was further relay imaged to the pulse compressor that was adjusted for optimal signal compression. Figure 3(a) presents a comparison of the measured cross-correlation traces for the compressed scattering noise and signal pulses. Being very different from PSF, the scattering noise stemming from chirped signal pulses can be compressed very well. The relatively lower temporal contrast was found to be related to the loss of spectral components when the conical ring propagated through the finite-size compressor. The spectrum of conical ring measured after the compressor, as given in the inset of Fig. 3(a), indicates that only one third of the spectrum was preserved. Figures 3(b) and 3(c) present the temporal profiles of the compressed signal pulses and scattering noise measured with FROG, respectively. The scattering noise had a compressed duration ~3 times that of the signal pulses.

Figure 4 shows the output energies of scattering noise and PSF as a function of the parametric gain. All the energy measurements in this paper were performed by averaging over 500 laser shots. The measurement of scattering noise was conducted repeatedly for incident signal energies of 90 $\mu$J and 9 $\mu$J. The results indicate that the energy of amplified scattering noise is proportional to both the incident signal energy and the parametric gain. We can thus deduce that the incident signal of 90 $\mu$J gives rise to an initial scattering noise of ~0.35 nJ that can be effectively amplified in the OPA stage, indicating an effective scattering ratio of $4\times10^{-6}$. This result implies that the scattering noise can be much stronger than the PSF that originates from quantum noise (half photon per mode). In our current setup, the scattering noise is stronger than PSF by 1-2 orders of magnitude. It is also worth mentioning that, as the scattering noise energy is proportional to the incident signal energy, the presence of scattering might question the effectiveness of the low-noise OPCPA design based on strong seeding [7].

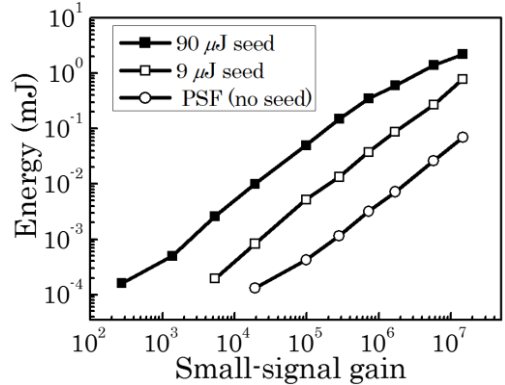

Fig. 4. Measured energies of the amplified scattering noise and PSF versus the small-signal gain of the OPA stage. The gain range was obtained by adjusting the pump pulse energy.

Finally, we study the scattering noise in a standard non-collinear phase-matching configuration that supports broadband signal amplification. The BBO crystal was tuned to $\theta = 23.8°$ and the seed signal was injected along a noncollinear angle of 4.0° relative to the pump beam. For this implementation, the scattering noise is amplified in parallel with the incident signal. Figure 5(a) presents the energy evolution of amplified signal and PSF with the increase of pump energy. The OPCPA output at the signal wavelength can reach an efficiency of 25.6% in the strong pump regime. Although the amount of PSF was negligible (always <

1% of the signal energy) in our single-stage OPCPA setup, we found that the OPCPA output contained considerable amount of scattering noise. A direct measurement of the amplified scattering noise energy is difficult due to its overlap with the amplified signal beam in space. In an alternative way, we study the propagation loss of OPCPA output energy. As illustrated by Fig. 5(b), at a pump pulse energy of 51 mJ ($I_p$= 2 GW/cm$^2$), the OPCPA output energy, defined in the aperture twice the signal beam diameter, decreased by 3.2% after a propagation distance of 2 m. With a stronger pump pulse of 88 mJ ($I_p$ = 3.4 GW/cm$^2$), this energy loss increased substantially to 12%. It suggests that the amplified scattering noise accounts for 12% of the total OPCPA output. Such a dramatic enhancement of scattering noise from an initial proportion (relative to the signal) of 4×10$^{-6}$ to a proportion of ~12% results from two main factors. Firstly, the noise gain generally exceeds signal gain due to a favorable phase-matching condition. Then, the saturation of signal amplification further contributes to a preferential amplification of the noise.

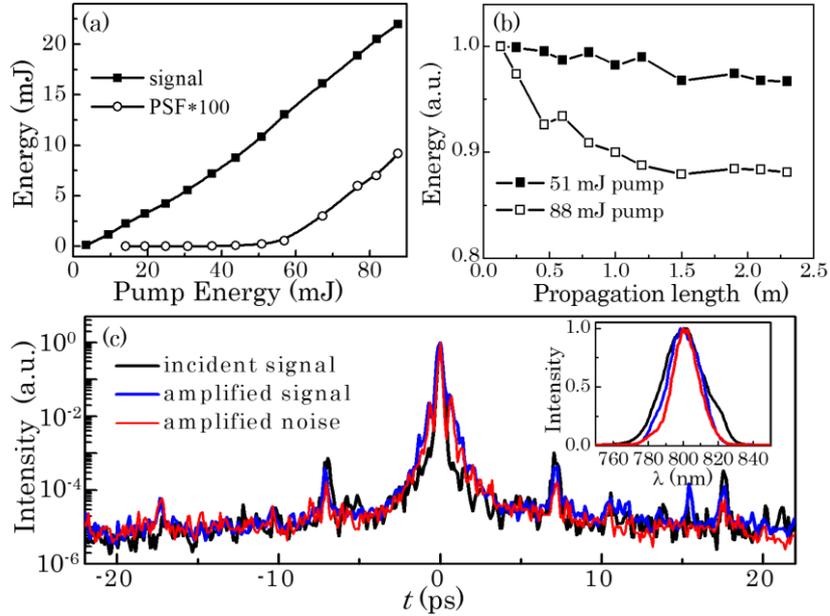

Fig. 5. (Color online) (a) Energies of amplified signal and PSF as a function of pump energy. The PSF energies had been multiplied by 100 to be readable. (b) The decrease of OPCPA output energy with propagation, measured for pump energies of 51 and 88 mJ, respectively. (c) The cross- correlation traces of the incident signal pulses (black), amplified signal pulses (blue) and amplified scattering noise (red) after compression. Inset: the corresponding spectra measured after the compressor.

Because the noncollinear phase-matching configuration supports a sufficiently large gain bandwidth (>100 nm), the amplified scattering noise was free from angular dispersion. We picked out a portion of amplified scattering noise in the focal plane of the relay telescope R$_3$ to study its temporal characteristics. Figure 5(c) gives the cross- correlation trace and spectrum of the compressed scattering noise, in comparison with that of the incident and amplified signal pulses. It indicates that the amplified scattering noise in this case can pass through the compressor without spectral loss. Consistently, there are also no observable differences in both pulse duration and contrast between the scattering noise and the signal pulses. This result suggests that the scattering noise has little influence on pulse contrast. In other words, the scattering noise is basically a kind of spatial noise in parametric amplifiers. This might be the reason why scattering noise was not perceived in the previous studies on OPCPA.

In conclusion, we have experimentally demonstrated that light scattering inherent in the signal beam path will contribute a non-ignorable noise term to the parametric amplifiers. In our OPCPA setup, approximately $4\times10^{-6}$ of the incident signal beam transfers to an initial scattering noise that can be further amplified in competition with the signal. Such scattering noise is typically much stronger than PSF. Even in a single-stage OPCPA as like our experimental situation, the amplified scattering noise can be as strong as 12% of the signal, while the PSF is negligible. We anticipate that the issue of scattering noise would be much more severe in multi-stage OPCPA systems. Our studies suggest that scattering noise is a crucial issue limiting extractable signal energy that should be considered in the design of OPCPA systems.

The work was partially supported by grants from the National Basic Research Program of China (973 Program) (2013CBA01505), the Natural Science Foundation of China (11121504) and the China Postdoctoral Science Foundation (2014M560332).